# Phase shift's influence of two strong pulsed laser waves on effective interaction of electrons

## S S Starodub[1], S P Roshchupkin[2] and V V Dubov[2]

[1]Institute of Applied Physics, National Academy of Sciences of Ukraine, 58 Petropavlovskaya Str., Sumy 40000, Ukraine

[2]Department of Theoretical Physics, Peter the Great St. Petersburg Polytechnic University, 195251, St-Petersburg, Russian Federation, Russia

E-mail: starodubss@gmail.com, serg9rsp@gmail.com, maildu@mail.ru

**Abstract**
The phase shift's influence of two strong pulsed laser waves on effective interaction of electrons was studied. Considerable amplification of electrons repulsion in the certain range of phase shifts and waves intensities is shown. That leads to electrons scatter on greater distances than without an external field. The value of the distance can be greater on 2-3 order of magnitude. Also considerable influence of the phase shift of pulses of waves on the possibility of effective attraction of electrons is shown.

Keywords: nonrelativistic electrons, strong femtosecond laser pulsed fields, effective interaction, phase shift.

## 1. Introduction

There are many works devoted to research of interaction of electrons in the presence of an electromagnetic field (see, the works [1-9]). The possibility of electron attraction in the presence of a plane electromagnetic wave was firstly shown by Oleinik [4]. However, the theoretical proof of the attraction possibility was given by Kazantsev and Sokolov for interaction of classical relativistic electrons in the field of a plane wave [5]. We also note the paper [6]. It is very important to point out, that attraction of classical electrons in the field of a plane monochromatic electromagnetic wave is possible only for particles with relativistic energies. In the authors works (see, review [3], articles [7-9]) the possibility of attraction of nonrelativistic electrons (identically charged ions) in the pulsed laser field was shown. Thus, in the review [3] the following processes were discussed: interaction of electrons (light ions) in the pulsed field of a single laser wave; interaction of nonrelativistic electrons in the pulsed field of two counter-propagating laser waves moving perpendicularly to the initial direction of electrons motion; the interaction of nonrelativistic light ions moving almost parallel to each other in the propagation direction of the pulsed field of two counter-propagating laser waves moving in parallel direction to ions; interaction of two nonrelativistic heavy nuclei (uranium 235), moving towards each other perpendicularly to the propagation direction of two counter-propagating laser waves. The effective force of interaction of two hydrogen atoms (after their ionization) in the pulsed field of two counter-propagating laser waves was considered in [7]. Influence of pulsed field of two co-propagating laser waves on the effective force of interaction of two electrons and two identically charged heavy nuclei was studied in [8]. The main attention is focused on the study of the influence of phase shifts of the pulse peak of the second wave relatively to the first on the effective force of particles interaction. The phase shift allows to increase duration of electron's confinement at a certain averaged effective distance by 1,5 time in comparison with the case of one and two counter-propagating pulsed laser waves. Interaction of two classical nonrelativistic electrons in the strong pulsed laser field of two light mutually perpendicular waves, when the maxima laser pulses coincide, was studied in [9]. It is shown that the effective force of electron interaction becoming the attraction force or anomalous repulsion force after approach of electrons to the minimum distance.

In the present work, in contrast to the mentioned above, interaction of two classical nonrelativistic electrons in the strong pulsed laser field of two light mutually perpendicular waves with the phase shifts of pulse peaks of the first and second waves is studied. It is shown that phase shifts of pulse peaks allow essentially change effective interaction of electrons than without phase shifts when the maxima laser pulses coincide (see, [3], [7], [9]). The obtained results can be used for experiments in the framework of modern research projects, where the sources of pulsed laser radiation are used (SLAC, FAIR) [10-12].

## 2. Equations of electron interaction in pulsed field of two laser waves

We study interaction of two nonrelativistic electrons moving towards each other along the axis $x$ in the field of two linearly polarized pulsed electromagnetic waves. Waves propagate perpendicularly to each other. The first wave propagates along the axis $z$, the second wave propagates along the axis $x$ (see figure 1).

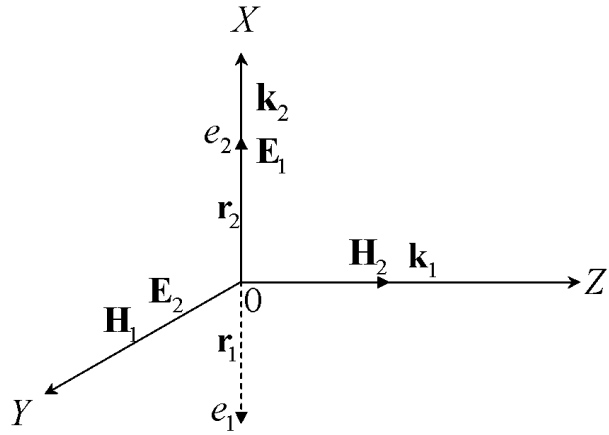

**Figure 1.** Interaction kinematics of two classical electrons in the field of two light mutually perpendicular waves.

We assume the strength of the electric and magnetic field in the following form:

$$\mathbf{E}(t, z_j, x_j) = \mathbf{E}_1(t, z_j) + \mathbf{E}_2(t, x_j), \tag{1}$$

$$\mathbf{E}_1(t, z_j) = E_{01} \cdot \exp\left[-\left(\frac{\varphi_{1j} - \delta\tau_1}{\omega_1 t_1}\right)^2\right] \cos\varphi_{1j} \cdot \mathbf{e}_x, \quad \varphi_{1j} = (\omega_1 t - k_1 z_j), \tag{2}$$

$$\mathbf{E}_2(t, x_j) = E_{02} \cdot \exp\left[-\left(\frac{\varphi_{2j} - \delta\tau_2}{\omega_2 t_2}\right)^2\right] \cos\varphi_{2j} \cdot \mathbf{e}_y, \quad \varphi_{2j} = (\omega_2 t - k_2 x_j), \tag{3}$$

$$\mathbf{H}(t, z_j, x_j) = \mathbf{H}_1(t, z_j) + \mathbf{H}_2(t, x_j), \tag{4}$$

$$\mathbf{H}_1(t, z_j) = H_{01} \cdot \exp\left[-\left(\frac{\varphi_{1j} - \delta\tau_1}{\omega_1 t_1}\right)^2\right] \cos\varphi_{1j} \cdot \mathbf{e}_y, \tag{5}$$

$$\mathbf{H}_2(t, x_j) = H_{02} \cdot \exp\left[-\left(\frac{\varphi_{2j} - \delta\tau_2}{\omega_2 t_2}\right)^2\right] \cos\varphi_{2j} \cdot \mathbf{e}_z, \tag{6}$$

where $\varphi_{ij}$ are phases of the corresponding wave ($i=1,2$) and corresponding electron ($j=1,2$); $E_{0i}$ and $H_{0i}$ are the strength of the electric and magnetic field in the pulse peak, respectively; $\delta\tau_i$ are phase shifts of pulse peaks of the first and second waves; $t_i$ and $\omega_i$ are the pulse duration and frequency of the first and the second wave; $\mathbf{e}_x$, $\mathbf{e}_y$, $\mathbf{e}_z$ are unit vectors directed along the $x$, $y$ and $z$ axes.

It is well-known fact that in the frame of the dipole approximation ($k=0$) and without taking into account the terms of the order $v/c \ll 1$ the particle interaction with the plane-wave field does not affect on the particle relative motion in the center-of-mass system. Thereby, we consider particle motion in the laser field beyond to the dipole approximation and an accuracy of quantities of the order $v/c \ll 1$ ($v$ is the relative velocity). Newton equations for motion of two identically charged particles with the mass $m$ and charge $e$ ($e=e_1=e_2$) in the pulsed field of two mutually perpendicular laser waves (1) - (6) are determined by the following expressions:

$$m\ddot{\mathbf{r}}_1 = -|e|\left[\mathbf{E}(t,z_1,x_1) + \frac{1}{c}\dot{\mathbf{r}}_1 \times \mathbf{H}(t,z_1,x_1)\right] - \frac{e^2}{|\mathbf{r}_2-\mathbf{r}_1|^3}(\mathbf{r}_2-\mathbf{r}_1), \qquad (7)$$

$$m\ddot{\mathbf{r}}_2 = -|e|\left[\mathbf{E}(t,z_2,x_2) + \frac{1}{c}\dot{\mathbf{r}}_2 \times \mathbf{H}(t,z_2,x_2)\right] + \frac{e^2}{|\mathbf{r}_2-\mathbf{r}_1|^3}(\mathbf{r}_2-\mathbf{r}_1), \qquad (8)$$

where $\mathbf{r}_1$ and $\mathbf{r}_2$ are electron radius-vectors.

Hereafter, the wave frequencies are the same: $\omega_1 = \omega_2 = \omega$, $|\mathbf{k}_1| = |\mathbf{k}_2| = k = \omega/c = \lambdabar^{-1}$.

Subsequent consideration we carry out in the center-of-mass system:

$$\mathbf{r} = \mathbf{r}_2 - \mathbf{r}_1. \qquad (9)$$

There are the following equations for particle relative motion in the center-of-mass system:

$$m\ddot{\mathbf{r}} = \frac{2e^2}{|\mathbf{r}|^3}\mathbf{r} - |e|(M_x\mathbf{e}_x + M_y\mathbf{e}_y + M_z\mathbf{e}_z), \qquad (10)$$

$$\begin{cases} M_x = f_1'\left[2\sin(\omega t)\sin\left(k\frac{r_z}{2}\right) - \frac{1}{c}\dot{r}_z\cos(\omega t)\right] + \frac{1}{c}f_2'\dot{r}_y\cos(\omega t), \\ M_y = f_2'\left[2\sin(\omega t)\sin\left(k\frac{r_x}{2}\right) - \frac{1}{c}\dot{r}_x\cos(\omega t)\right], \\ M_z = \frac{1}{c}f_1'\dot{r}_x\cos(\omega t), \end{cases} \qquad (11)$$

$$f_1' = E_{01}\exp\left[-\frac{(\omega t - \delta\tau_1)^2}{(\omega t_1)^2}\right], \quad f_2' = E_{02}\exp\left[-\frac{(\omega t - \delta\tau_2)^2}{(\omega t_2)^2}\right]. \qquad (12)$$

Note that equations for relative motion (10), (11) are written beyond to the dipole approximation ($k \neq 0$) and an accuracy of term of order $v/c \ll 1$. Note, small influence of external strong pulsed laser field on radius-vector of the center-of-mass motion was shown in the work [9]. Thereby, study of relative motion of electrons makes sense to be done only.

Equations (10) - (11) can be written in the dimensionless form:

$$\ddot{\boldsymbol{\xi}} = \mathbf{F}, \quad \mathbf{F} = \beta \cdot \frac{\boldsymbol{\xi}}{|\boldsymbol{\xi}|^3} - (N_x\mathbf{e}_x + N_y\mathbf{e}_y + N_z\mathbf{e}_z), \qquad (13)$$

$$\begin{cases} N_x = \eta_1 f_1 \left[ \sin(\tau) \sin\left(\frac{\xi_z}{2}\right) - \frac{\dot{\xi}_z}{2} \cos(\tau) \right] + \eta_2 f_2 \dot{\xi}_y \cos(\tau), \\ N_y = \eta_2 f_2 \left[ \sin(\tau) \sin\left(\frac{\xi_x}{2}\right) - \frac{\dot{\xi}_x}{2} \cos(\tau) \right], \\ N_z = \eta_1 f_1 \frac{\dot{\xi}_x}{2} \cos(\tau), \end{cases} \quad (14)$$

where,

$$\xi = k\mathbf{r} = \mathbf{r}/\lambdabar, \quad \tau = \omega t, \quad \tau_{1,2} = \omega t_{1,2}, \quad (15)$$

$$\eta_i = \frac{|e|E_{0i}}{\mu c \omega}, \quad \beta = \frac{e^2/\lambdabar}{\mu c^2}; \quad \mu = m/2, \quad f_i = \exp\left(-\frac{(\tau - \delta\tau_i)^2}{\tau_i^2}\right), \quad i = 1, 2. \quad (16)$$

Here, $\xi$ is the radius-vector of the relative distance between electrons in unit of the wavelength, the parameters $\eta_{1,2}$ are numerically equal to the ratio of the oscillation velocity of an electron in the peak of a pulse of the first or second wave to the velocity of light $c$ (hereinafter, we consider parameters $\eta_{1,2}$ as oscillation velocities); the parameter $\beta$ is numerically equal to the ratio of the energy of Coulomb interaction of electrons with the reduced mass $\mu$ at the wavelength to the particle rest energy.

The pulse duration exceeds considerably the period of wave rapid oscillation ($\sim \omega^{-1}$) for a majority of modern pulsed lasers:

$$\tau_{1,2} \gg 1. \quad (17)$$

Consequently, the relative distance between electrons should be averaged over the period of wave rapid oscillation:

$$\bar{\xi} = \frac{1}{2\pi} \int_0^{2\pi} \xi \cdot d\tau. \quad (18)$$

We note that expressions (13), (14) consider interaction with the Coulomb field and the pulsed-wave field strictly, and don't have the analytical solution. For subsequent analysis we will study all equations numerically.

Electrons initial relative coordinates and velocities are the following:

$$\begin{aligned} \xi_{x0} = 2, \quad \xi_{y0} = 0, \quad \xi_{z0} = 0, \\ \dot{\xi}_{x0} = -1.7 \cdot 10^{-3}, \quad \dot{\xi}_{y0} = 0, \quad \dot{\xi}_{z0} = 0. \end{aligned} \quad (19)$$

The interaction time is $\tau = 1200$ ($\tau \in [-600 \div 600]$), ($t = 600\,\text{fs}$) and it was increased, if necessary for more clear results. Frequencies of waves are $\omega_1 = \omega_2 = 2\,\text{Ps}^{-1}$ ($\lambdabar = 0.15\,\mu\text{m}$), pulse durations are $\tau_1 = \tau_2 = 600$ ($t_1 = t_2 = 300\,\text{fs}$). Field intensities (oscillations velocities $\eta_1, \eta_2$) are varied. Phase shifts are vary within $\delta\tau_{1,2} \in [-600 \div 600]$ and step is $h = 50$. Initial conditions are the same as in [9]. That allows to estimate influence of phase shifts on relative motion of electrons and compare results. Note, in the work [9] the parameter of the phase shift of a pulse of a wave ($\delta\tau_{1,2} = 0$) was absent, and pulse peaks of both waves were in moment $\tau = 0$. Initial coordinates and velocities of electrons are chosen so that at the point $\tau = 0$ electrons were in maximum approach (the Coulomb force was maximum). In this work the pulse peaks of waves can have maximum at any moment of time ($\tau = \delta\tau_{1,2}$) and it's leads to significant change in the behavior of electron interaction. Numerical solving of equations for relative motion (13) results to several cases.

## 3. Anomalous repulsion of electrons

The case when the oscillation velocity of the first wave is greater than the initial velocity of electrons ($\eta_1 > \dot{\xi}_0$), and oscillation velocity of the second wave considerably exceeds the initial velocity ($\eta_2 \gg \dot{\xi}_0$). Calculations over all values of phase shifts allowed to find out areas of anomalous repulsion of electrons. In this areas electrons can scatter at very long distances exceeding the distance of electron scattering without an external field in hundreds of times (see figures 2 (a), (b)). Let designate the final distance at which electrons scatter in the time moment as $\tau_{Cfinal} = 600$: without an external field as $\bar{\xi}_{Cfinal} = 2$; in the external field, when $\delta\tau_1 = 0$, $\delta\tau_2 = 0$ as $\bar{\xi}^{(0)}_{final}$; in an external field, when $\delta\tau_1 \neq 0$, $\delta\tau_2 \neq 0$ as $\bar{\xi}_{final}$.

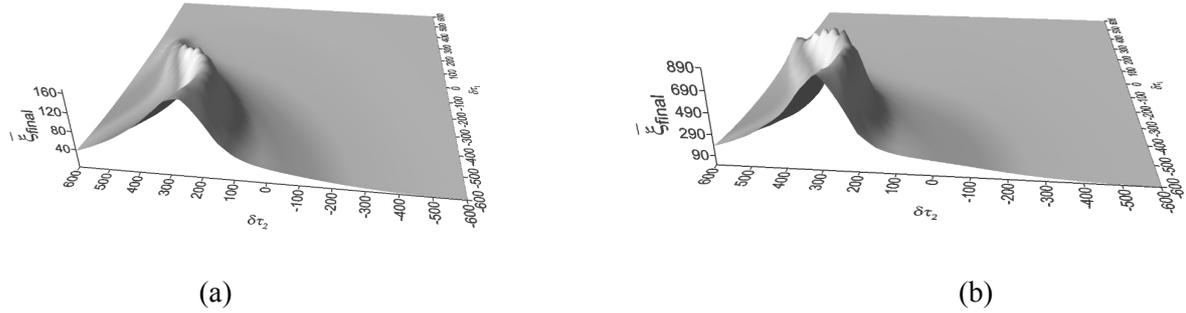

(a)                                                           (b)

**Figure 2.** The final averaged relative distance $\bar{\xi}_{final}$ (in the time moment $\tau_{Cfinal} = 600$, $\bar{\xi}_{Cfinal} = 2$) against different phase shifts of pulse peaks $\delta\tau_1$, $\delta\tau_2$. Oscillation velocities: $\eta_1 = 3 \times 10^{-3}$, (a) - $\eta_2 = 6 \times 10^{-2}$, (b) - $\eta_2 = 10^{-1}$ (field intensities: $I_1 = 3.4 \times 10^{12}$ W/cm$^2$, (a) - $I_2 = 1.3 \times 10^{15}$ W/cm$^2$, (b) - $I_2 = 3.8 \times 10^{15}$ W/cm$^2$).

Figures 2 (a, b) show dependence of the final distance $\bar{\xi}_{final}$ at which the electrons scatter at the time moment $\tau_{Cfinal} = 600$ for different values of phase shifts $\delta\tau_1$, $\delta\tau_2$ and for two values of the oscillation velocity of the second wave. One can see, that the final distance $\bar{\xi}_{final}$ is considerably depends from the oscillation velocity of the second wave and have maximum values for next ranges of the phase shifts: $\delta\tau_1 \in [-600 \div -50]$, $\delta\tau_2 \in [100 \div 500]$. Thus, for the oscillation velocity $\eta_2 = 6 \times 10^{-2}$ the final distance can reach the value $\bar{\xi}_{final} \approx 160$ (see figure 2 a), and for the oscillation velocity $\eta_2 = 10^{-1}$ the final distance can reach the value $\bar{\xi}_{final} \approx 900$ (see figure 2 b). Figures 3, 4 show dependence of averaged relative distance $\bar{\xi}$ against the interaction time $\tau$ for mainly interesting values of the phase shifts. It's seen that taking into account of the phase shifts of pulse peaks can considerably increase the repulsion force. Thus, for $\eta_1 = 3 \times 10^{-3}$, $\eta_2 = 6 \times 10^{-2}$ and $\delta\tau_1 = -550$, $\delta\tau_2 = 250$ ratio $\bar{\xi}_{final}/\bar{\xi}_{Cfinal} \approx 80$ and ratio $\bar{\xi}_{final}/\bar{\xi}^{(0)}_{final} \approx 32$ (see the curve 3 on figure 3), and for oscillation velocities $\eta_1 = 3 \times 10^{-3}$, $\eta_2 = 10^{-1}$ and $\delta\tau_1 = -450$, $\delta\tau_2 = 250$ the ratio $\bar{\xi}_{final}/\bar{\xi}_{Cfinal} \approx 450$ and the ratio $\bar{\xi}_{final}/\bar{\xi}^{(0)}_{final} \approx 180$ (see the curve 3 on figure 4).

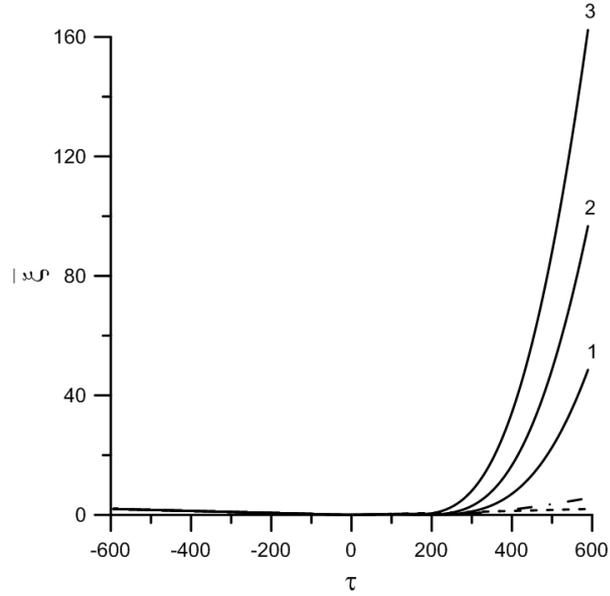

**Figure 3.** The averaged relative distance $\bar{\xi}$ against the interaction time $\tau$. The dashed line corresponds to the case of the absence of the external field. The dashed-dot line and solid lines correspond to oscillation velocities: $\eta_1 = 3 \times 10^{-3}$, $\eta_2 = 6 \times 10^{-2}$ (field intensities: $I_1 = 3.4 \times 10^{12}\ \text{W}/\text{cm}^2$, $I_2 = 1.3 \times 10^{15}\ \text{W}/\text{cm}^2$), phase shifts of pulse peaks: 1- $\delta\tau_1 = -450$, $\delta\tau_2 = 450$; 2- $\delta\tau_1 = -350$, $\delta\tau_2 = 300$; 3- $\delta\tau_1 = -550$, $\delta\tau_2 = 250$; the dashed line with a dot - $\delta\tau_1 = 0$, $\delta\tau_2 = 0$.

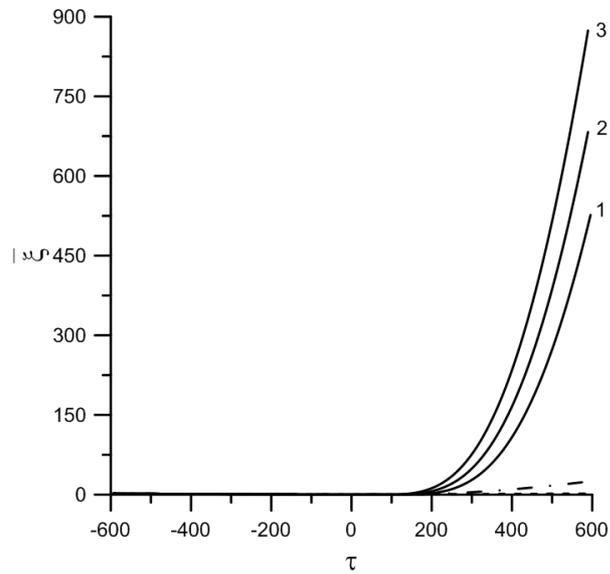

**Figure 4.** The averaged relative distance $\bar{\xi}$ against the interaction time $\tau$. The dashed line corresponds to case without external field. The dashed-dot line and solid lines correspond to oscillation velocities: $\eta_1 = 3 \times 10^{-3}$, $\eta_2 = 10^{-1}$ (the field intensities: $I_1 = 3.4 \times 10^{12}\ \text{W}/\text{cm}^2$, $I_2 = 3.8 \times 10^{15}\ \text{W}/\text{cm}^2$), the phase shifts of pulse peaks: 1- $\delta\tau_1 = -300$, $\delta\tau_2 = 350$; 2- $\delta\tau_1 = -400$, $\delta\tau_2 = 300$; 3- $\delta\tau_1 = -450$, $\delta\tau_2 = 250$; the dashed-dot line - $\delta\tau_1 = 0$, $\delta\tau_2 = 0$.

## 4. The effective slowing-down of electrons

The case, when the oscillation velocity $\eta_1$ has to be close to the initial relative velocity $\eta_1 \approx \dot{\xi}_0$ and the oscillation velocity $\eta_2$ is greater an order of magnitude. Increasing of the interaction time allows us to see oscillations of the effective attraction of electrons. Electrons, after approaching and scattering, get the strong pulse of the attraction and then they re-approach. Let designate the time at which the averaged relative distance between electrons is equal to $\bar{\xi}_{Cfinal} = 2$: in the external field, when $\delta\tau_1 = 0$, $\delta\tau_2 = 0$ - $\tau_{final}^{(0)}$; in the external field, when $\delta\tau_1 \neq 0$, $\delta\tau_2 \neq 0$ - $\tau_{final}$. Figure 5 shows dependence of the final distance $\bar{\xi}_{final}$ at which electrons scatter at the time $\tau = 2000$ for different values of phase shifts $\delta\tau_1$, $\delta\tau_2$ and next values of oscillation velocities $\eta_1 = 1.7 \times 10^{-3}$, $\eta_2 = 3 \times 10^{-2}$. It's seen that the final distance $\bar{\xi}_{final}$ is smaller $\bar{\xi}_{Cfinal} = 2$ and it has the minimum value down to $\bar{\xi}_{final} = 10^{-1}$ for the next ranges of phase shifts $\delta\tau_1 \in [200 \div 400]$, $\delta\tau_2 \in [100 \div 600]$.

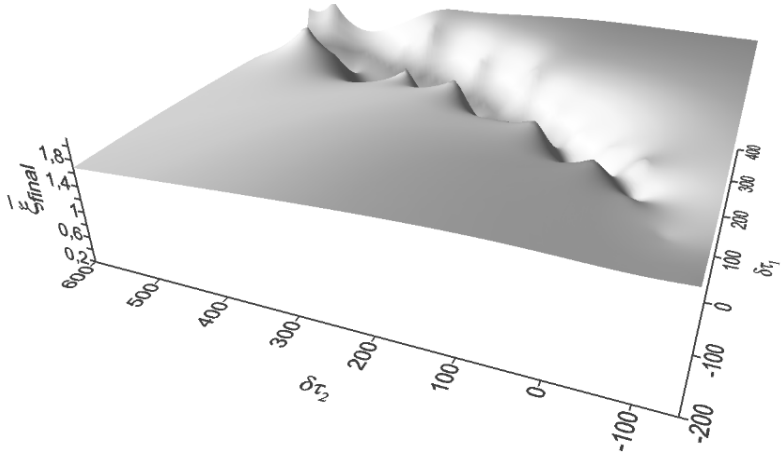

**Figure 5.** The final averaged relative distance $\bar{\xi}_{final}$ (in the time moment $\tau = 2000$, $\bar{\xi}_{Cfinal} = 2$) against different phase shifts of the pulse peaks $\delta\tau_1$, $\delta\tau_2$. Oscillation velocities: $\eta_1 = \dot{\xi}_0 = 1.7 \times 10^{-3}$, $\eta_2 = 3 \times 10^{-2}$ (field intensities: $I_1 = 1.1 \times 10^{12}$ W/cm$^2$, $I_2 = 3.4 \times 10^{14}$ W/cm$^2$).

Figure 6 shows dependence of averaged relative distance $\bar{\xi}$ on the interaction time $\tau$ for mainly interesting values of the phase shifts. One can see that taking into account of the phase shifts can considerably increase the attraction force. Thus, the time of electron scattering to the initial value of the distance ($\bar{\xi}_{Cfinal} = 2$) is increased in comparison with the case without an external field to $\tau_{final}/\tau_{Cfinal} \approx 13.5$ (see figure 6 curve 1 and the dashed line); in the external field when $\delta\tau_1 = 0$, $\delta\tau_2 = 0$ the time is increased to $\tau_{final}/\tau_{final}^{(0)} \approx 4$ (see figure 6 curve 1 and the dashed-dot line) for oscillation velocities $\eta_1 = 1.7 \times 10^{-3}$, $\eta_2 = 3 \times 10^{-2}$ and phase shifts $\delta\tau_1 = 350$, $\delta\tau_2 = 400$. The effect is a bit weaker for phase shifts $\delta\tau_1 = 300$, $\delta\tau_2 = 300$ (see figure 6 curve 2).

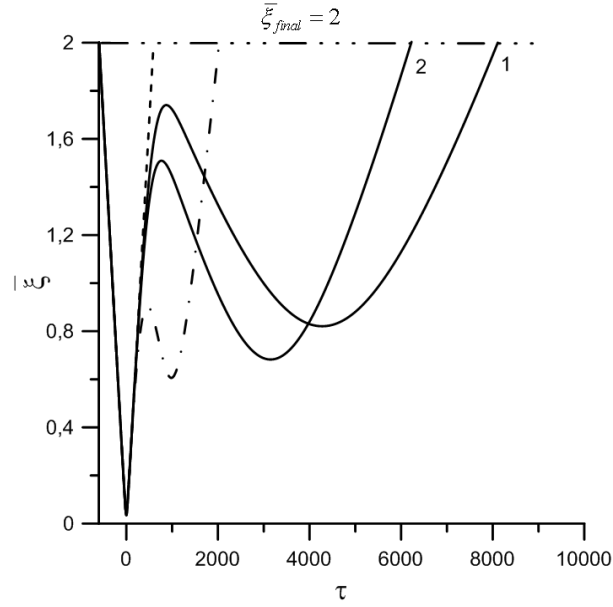

**Figure 6.** The averaged relative distance $\bar{\xi}$ against the interaction time $\tau$. The dashed line corresponds to case without external field. The dashed-dot line and solid lines correspond to the oscillation velocity: $\eta_1 = \dot{\xi}_0 = 1.7 \times 10^{-3}$, $\eta_2 = 3 \times 10^{-2}$ (the field intensity: $I_1 = 1.1 \times 10^{12}$ W/cm$^2$, $I_2 = 3.4 \times 10^{14}$ W/cm$^2$), phase shifts of pulse peaks: 1- $\delta\tau_1 = 350$, $\delta\tau_2 = 400$, 2- $\delta\tau_1 = 300$, $\delta\tau_2 = 300$; the dashed-dot line - $\delta\tau_1 = 0$, $\delta\tau_2 = 0$.

## 5. Conclusion

Performed study shows that taking into account of phase shifts of pulse peaks can essentially change the behavior of the effective interaction of electrons:

1. The anomalous repulsion of electrons is observed when the oscillation velocity of the first wave is greater than the initial velocity of electrons ($\eta_1 > \dot{\xi}_0$), and the oscillation velocity of the second wave is considerably greater ($\eta_2 \gg \dot{\xi}_0$). Thus, the maximum effect of anomalous repulsion of electrons corresponds to the following ranges of phase shifts of pulse peaks: $\delta\tau_1 \in [-300 \div -550]$, $\delta\tau_2 \in [250 \div 450]$. So, for intensities of the waves $I_1 = 3.4 \times 10^{12}$ W/cm$^2$, $I_2 = 3.8 \times 10^{15}$ W/cm$^2$ and phase shifts $\delta\tau_1 = -450$, $\delta\tau_2 = 250$ the ratio $\bar{\xi}_{final}/\bar{\xi}_{Cfinal} \approx 450$, and the ratio $\bar{\xi}_{final}/\bar{\xi}_{final}^{(0)} \approx 180$.

2. The effective attraction of electrons takes place, when the oscillation velocity $\eta_1$ has to be close to the initial relative velocity ($\eta_1 \approx \dot{\xi}_0$), and the oscillation velocity of the second wave is greater in an order of the magnitude. Thus, the maximum effect of slowing-down of electrons corresponds to the following ranges of phase shifts of pulse peaks: $\delta\tau_1 \in [350 \div 400]$, $\delta\tau_2 \in [400 \div 550]$. So, for intensities of waves $I_1 = 1.1 \times 10^{12}$ W/cm$^2$, $I_2 = 3.4 \times 10^{14}$ W/cm$^2$ and phase shifts $\delta\tau_1 = 400$, $\delta\tau_2 = 550$ the values of slowing-down of electrons may be equal $\tau_{final}/\tau_{Cfinal} \approx 16.5$ and $\tau_{final}/\tau_{final}^{(0)} \approx 5$.

The obtained results can be used for experiments in the framework of modern research projects, where the sources of pulsed laser radiation are used (SLAC, FAIR) [10-12].